\documentclass{article}
\usepackage{sao2,psfig}
\issue{2000, 50, 5-38}
\newcommand{\E}{\rlap{E}}
\begin{document}
\markboth{Karachentsev et al.}{A list of peculiar velocities of RFGC galaxies}
\title{A list of peculiar velocities of  RFGC galaxies}
\author{I.D. Karachentsev \inst{a} \and V.E. Karachentseva \inst{b} \and
Yu.N. Kudrya \inst{b} \and D.I. Makarov \inst{a} \and S.L. Parnovsky
\inst{b}}
\institute{\saoname \and Astronomical Observatory of the Kyiv Taras
Shevchenko National University, Observatorna 3, 04053, Kyiv, Ukraine}
\date{August, 2, 2000}{September 27, 2000}
\maketitle \begin{abstract} A list of
radial velocities, HI line widths and peculiar velocities of 1327 galaxies
from the RFGC catalogue has been compiled using actual observations and
literature data. The list can be used for studying bulk motions of galaxies,
construction of the field of peculiar velocities and other tasks.
 \end{abstract}
\section{Introduction} The study of non-Hubble
motions and of the peculiar velocity field of galaxies on different scales forms
an observational foundation for many tasks of present-day cosmology (see the
survey of Strauss and Willick, 1995). For detailed analysis of peculiar
velocities ample samples of galaxies with measured radial velocities $V_h$ and
independent distance estimates $Hr$ are required. The first homogeneous
catalogue of this kind, Mark\,III, was compiled as a result of integration and
unification of data of observations of radial velocities and rotational
velocities (in the radio or optical ranges) of galaxies from different samples,
which were made by different groups of authors in 1982--1996 (Willick et al.,
1995, 1997b). It was based on mutual coordination of the
TF relations (Tully--Fisher, 1977),  which were represented in the form
``luminosity --- line width'' for the samples of spiral galaxies of clusters
and of the field. The sample of spirals (Han and Mould, 1992) was adopted as
the reference one. The $D_n-\sigma$ samples were later also included in
Mark\,III, and relative zero points for the arrays of E and S galaxies
(Dekel et al., 1999) were determined. Without going into details of
constructing Mark\,III, which were described comprehensively by the authors,
let us note that standardization of the selection criterion was an
important part of the work since the initial observational samples
incorporated galaxies from different optical catalogues with different
limits in apparent magnitude or angular diameter and having different angles
of inclination to the line of sight. Notice that strongly inclined, ``flat''
spirals were not originally included in Mark\,III. The total number of
individual galaxies involved in the final version of Mark\,III amounts to
3000; they were divided into 1200 groups. Thus, the catalogue Mark\,III
is important for it contains homogeneous data for early- and late-type galaxies,
both cluster members and field galaxies and distributed practically throughout
the sky. This allows a map of matter density distribution to be constructed
from the distribution of peculiar velocities, and the values of cosmological
parameters to be obtained.

There is another approach to establishing a homogeneous sample of galaxies
for studying the non-Hubble motions of galaxies. It was realized when
compiling the catalogue of flat edge-on galaxies, RFGC (Karachentsev et al.,
1999a; the first version is FGC, Karachentsev et al., 1993). The main idea
of this approach is a
special all-sky search for late edge-on spiral galaxies and
selection for the catalogue of objects satisfying the conditions $a/b\ge7$
and $a\ge0.6'$, where $a$ and $b$ are the major and minor axes. The RFGC
catalogue contains 4236 galaxies and covers the entire sky. Since the
selection was performed using the surveys POSS-I and ESO/SERC, which have
different photometric depth, we reduced the diameters of the southern-sky
galaxies to the system POSS-I, which turned out to be close to the system
$a_{25}$. The substantiation of selecting exactly flat galaxies and a
detailed analysis of optical properties of the catalogue objects
are available
in the texts of FGC, RFGC and in the papers of the authors cited therein.

Some extensive work has so far been done on measurements of radial velocities,
HI 21\,cm line widths, $W_{50}$, or rotational curves  $V_{rot}$ for the galaxies
of the RFGC catalogue. Besides, we have collected such data from literature.
Part of the data have been used to derive the radial velocities and
direction of bulk motion of flat galaxies (Karachentsev et al., 1995, 1999b).

Herein we report the radial velocities, HI line widths and peculiar
velocities for 1327 galaxies of the RFGC catalogue.

\section{Samples}
Observational data were divided into several samples.
\begin{enumerate}
\item The observations of flat galaxies from FGC performed with the 305\,m telescope at Arecibo
(Giovanelli et al., 1997). The observations are confined within the zone
$0\degr<\delta\le+38\degr$ accessible to the radio telescope. There was
no selection by the visible angular diameter, type, axes ratio and other
characteristics. We have not included in the summary the flat galaxies
from the Supplement to FGC, which do not satisfy the condition $a/b\ge7$,
and also the galaxies with uncertain values of $W_{50}$, in accordance
with the notes in the paper by Giovanelli et al. (1997). Our list contains
490 flat galaxies from this paper.
\item The observations of optical rotational curves made with the 6\,m
telescope of SAO RAS (Makarov et al., 1997\,a,\,b; 1999; 2000).
The objects located in the zone $\delta \ge 38\degr$,
with the axes ratio $a/b \ge 8$ and a large diameter $a\le 2'$ were selected
for the observations. The
maximum rotational velocities were converted to $W_{50}$ by a relation
derived through comparison of  optical and radio observations of 59
galaxies common with sample ``1'' (Makarov et al., 1997a). 300 galaxies from
these papers are included into our list.
\item The data on radial velocities and hydrogen line widths in the FGC
galaxies identified with the RC3 catalogue (de Vaucouleurs et al., 1991).
In a few cases, where only $W_{20}$ are available in RC3, they were converted
to $W_{50}$ according to Karachentsev et al. (1993). This sample comprises
flat galaxies all over the sky, a total of 167 objects.
\item The data on HI line widths (64\,m radio telescope, Parkes) and
on optical rotational curves $V_{rot}$ (2.3\,m telescope of Siding Spring)
for the flat galaxies identified with the lists by Mathewson et al. (1992),
Mathewson and Ford (1996). The optical data were converted to the widths
$W_{50}$ according to Mathewson and Ford (1996). The Sb--Sd galaxies from
the catalogue ESO/Uppsala (Lauberts, 1982) with angular dimensions
$a\ge 1'$, inclinations $i>40\degr$, and a galactic latitude $(|b|)\ge 11\degr$
have been included in the lists.  As Mathewson et al. (1992) report, the data
obtained with the 64\,m and 305\,m telescopes are in good agreement. Our sample
contains 177 flat galaxies from these papers.
\item The HI line observations of
flat galaxies carried out by Matthews and van Driel (2000) using the radio
telescopes in Nancay ($\delta > -38\degr$) and Green Bank ($\delta =-38\degr
\div -44.5\degr$). They have selected the flat galaxies from FGC(E) with angular
dimensions $a>1'$, of Scd types and later, mainly of low surface brightness
(SB\,=\,III and IV according to RFGC). We did not include in our list uncertain
measurements from the data of Matthews and van Driel (2000). In the case of
common objects with samples ``1'' or ``2'' we excluded the data of
Matthews and van Driel (2000) on the basis of the comparisons of $V_h$ and
$W_{50}$. The subsample ``5'' comprises 193 galaxies.
\end{enumerate}

 Thus, each galaxy enters into
our sample with a single set of $V_h$ and $W_{50}$ estimates. We have included
in the list a total of 1327 flat galaxies. The heliocentric velocities
were reduced to the system CMB, $V_{3K}$ according to Kogut et al. (1993),
and to the centroid of the Local Group, $V_{LG}$ (Karachentsev and Makarov,
1996). The observed widths were corrected for the cosmological broadening
and turbulence (Tully and Fouqu\'{e}, 1985). The angular diameters were
corrected for the extinction in the Galaxy and intrinsic absorption
(Karachentsev, 1991).

The mean characteristics of the galaxies are collected in Table\,1. The
distribution of the catalogue blue diameters as a function of $V_{3K}$
for 1327 flat galaxies is shown in Fig.\,1. It can be seen that the diameters of
galaxies are close to the limiting catalogue value, $a=0.6'$, and this
holds practically for all (but for the smallest) distances. Thus the selection
by angular diameters in our list is  not strong, although
individual samples have different depths (Table\,1) and are differently
located on the lg\,$a-V_{3K}$ diagram (we do not present  the figures here).

\begin{figure*}
\centerline{\psfig{figure=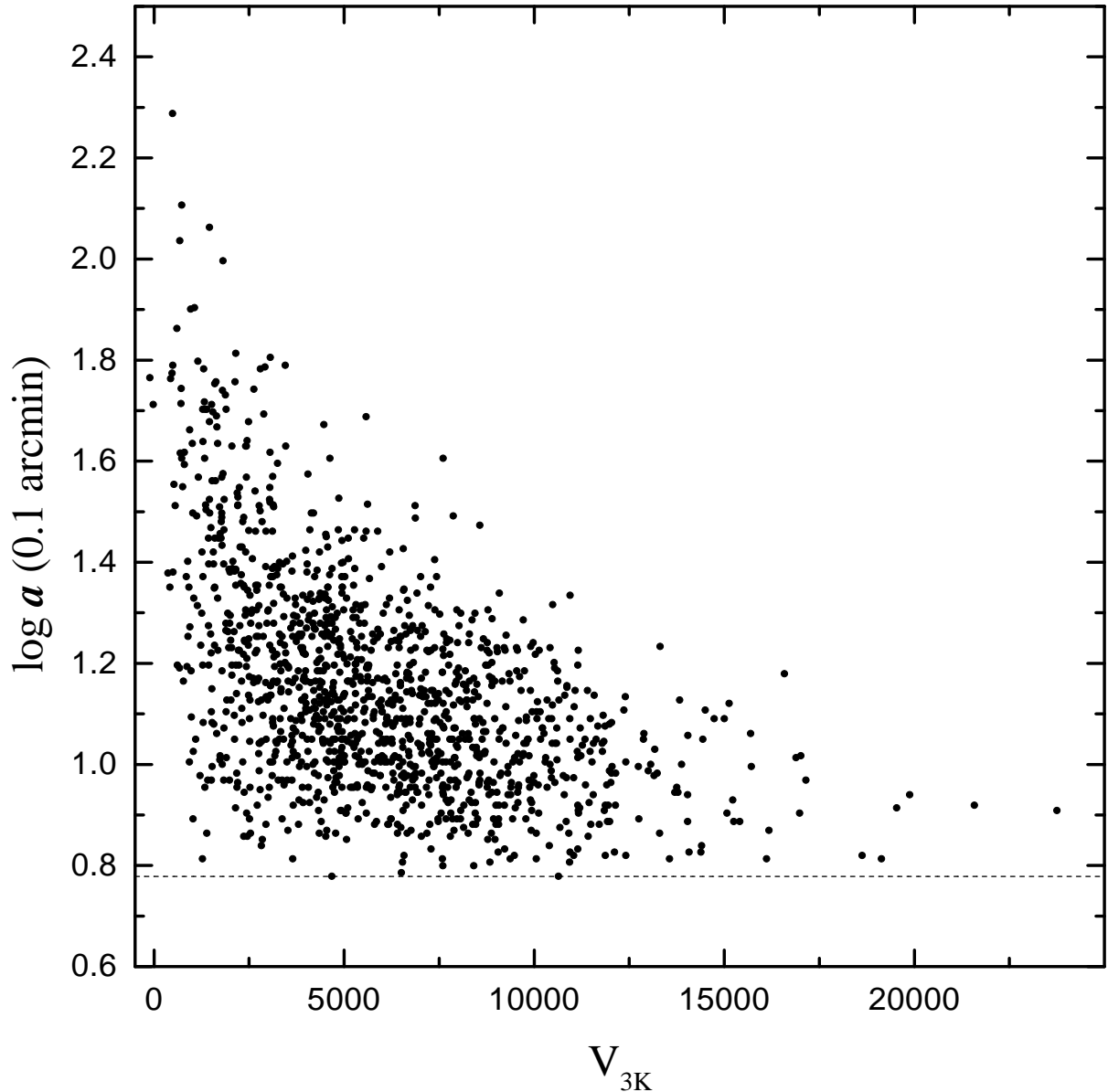,width=16cm}}
\caption{
The relationship between ``blue'' diameters of 1327 RFGC galaxies and
radial velocity $V_{3K}$.}
\end{figure*}

\begin{table}
\caption{Mean sample characteristics of flat galaxies}
\begin{center}
\begin{tabular}{ccccc}\hline
Sample & N & $<a>^{cor}_{b}$  & $<V_{3K}>$  & $<W^{cor}_{50}>$  \\
       &   &    arcsec           &    km/s     &   km/s          \\
\hline
1 & 490 & $1.02\pm 0.22$ & $6531\pm 54$ & $299\pm 5$ \\
2 & 300 & $0.81\pm 0.16$ & $8112\pm 62$ & $313\pm 6$ \\
3 & 167 & $2.28\pm 0.41$ & $3587\pm 50$ & $276\pm 10$ \\
4 & 177 & $1.40\pm 0.28$ & $4855\pm 50$ & $276\pm 8$ \\
5 & 193 & $0.80\pm 0.14$ & $4414\pm 48$ & $186\pm 6$ \\
All & 1327 & $1.15\pm 0.29$ & $5986\pm 58$ & $280\pm 3$ \\
\hline
\end{tabular}
\end{center}
\end{table}

The distribution of 1327 flat galaxies over the sky in the galactic coordinates
is displayed in Fig.\,2. One can see that they cover the sky fairly
uniformly except the Milky Way, with some excess in the number of galaxies
north of $\delta=0$.

\begin{figure*}
\centerline{\psfig{figure=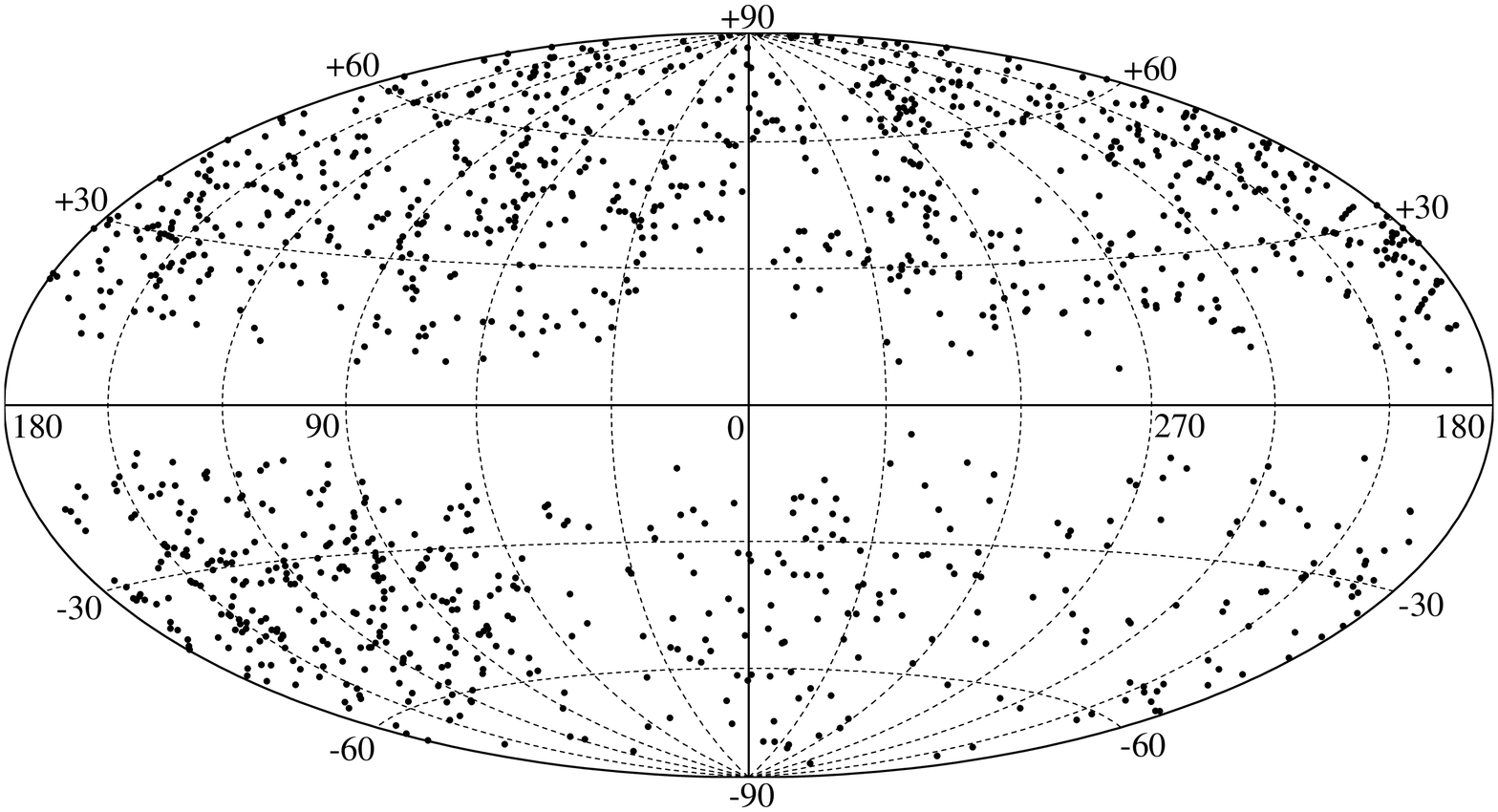,width=16cm}}
\caption{
The distribution of 1327 galaxies over the sky in galactic coordinates.}
\end{figure*}

\section{Computation of peculiar velocities}
Before  description of determination of the distance to a RFGC
galaxy independent upon the radial velocity, we will make some general remarks.
For each of the galaxies included
in our list, we have a homogeneous set of catalogue parameters: 1) measured
``blue'' and ``red'' angular diameters $a_b, b_b, a_r, b_r$ in a system
close to $a_{25}$; 2) Hubble type Ty (5
corresponds to type Sc); 3) surface brightness index SB (brightness
decreases from I to IV) etc. We used the TF relation in the form ``linear
diameter -- width'' and therefore, in distinction to the Mark\,III compilers,
did not need to coordinate apparent magnitudes observed in different
photometric bands and with different apertures.

The data on radial velocities and widths were taken from different lists,
and in the case of ``duplicates'' we took account of the data of only one
source. Each galaxy entered into our list with the equal weight. The
second principal distinction of our list from Mark\,III is that the distances
to galaxies were determined not by coordination of the TF relations for
individual samples, but from the generalized TF relation common to all
galaxies.

Karachentsev et al. (2000) describe in detail the procedure of determining the
distances to galaxies from the generalized Tully-Fisher relation. The
distance $R=Hr$ (expressed in km/s and corresponding to the radial velocity of
a galaxy in the case of purely Hubble expansion) was represented as a
linear combination of functions, dependent on the galaxy characteristics,
with coefficients $C_i$. Using the Fisher criterion the statistical
significance was determined for each regressor. The regressors, which
entered into the approximation $R$ with a confidence level of less than 99.9\%,
were rejected. We use in this paper the distance approximation

%$$
\begin{eqnarray}
R & = &(C_1+C_2B+C_3BT)W_{50}/a_r+C_4W_{50}/a_b \nonumber \\
& & {}+C_5(W_{50})^2/(a_r)^2+C_6/a_r.\nonumber
\end{eqnarray}
%$$

It is implied here  the corrected diameters and widths; $B=SB-2$, and
$T=Ty-5.35$. As compared to our paper of 2000, the summand, which included
the axes ratio, has been discarded in this approximation as statistically
insignificant, and a new statistically significant regressor $W_{50}/a_b$
is added.

Considering the model of bulk motion of galaxies to be the most simple dipole
approximation, we represented the measured radial velocity of each galaxy
$V_{3K}$ in the relic radiation rest frame as the sum of three components: R, the projection of the ordered
(bulk) motion,
$$ V_d=(V^{B}_{x}\cos{l}\cos{b}+V^{B}_{y}\sin{l}\cos{b}+V^{B}_{z}\sin{b}),$$
and a random small-scale component $\epsilon_V$. Using the least-square
method, we derived the $C_i$ coefficients of the generalized Tully-Fisher
relation and the values of $V^{B}_{x}, V^{B}_{y}, V^{B}_{z}$ (in km/s) with
which the sum of squares of the deviations $\epsilon^2$ is minimized. For
all the galaxies of the list peculiar velocities were computed
$$V_{pec}=V_{3K}-Hr-V_d.$$
Having excluded 56 galaxies with peculiar velocities
over $3\sigma$ and having introduced a restriction $R_{max}=10000$\,km/s to
make the sample more homogeneous in depth (see Fig.\,1), we derived a
generalized TF relation with coefficients:
$$ C_1=(18.1\pm1.6), C_2=(2.0\pm0.2),$$
$$ C_3=(-0.85\pm0.15), C_4=(6.8\pm1.4),$$
$$ C_5=(-7.6\pm1.2)\cdot10^{-3}, C_6=(-813\pm99),$$
$$ V^{B}_{x}=(261\pm68), V^{B}_{y}=(-212\pm71), V^{B}_{z}=(-7\pm56).$$

\begin{figure*}
\centerline{\psfig{figure=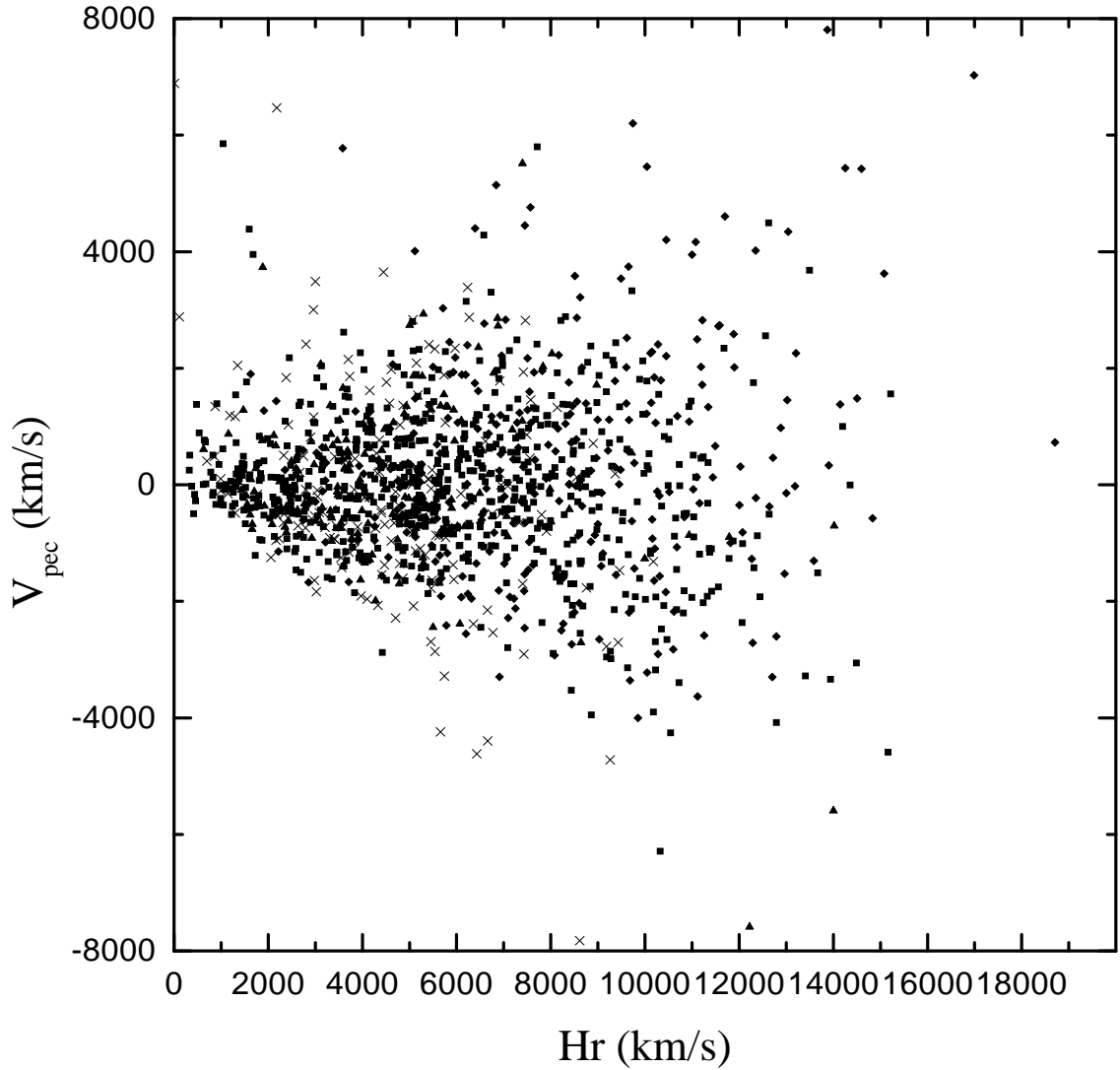,width=15cm}}
\caption{
The ``peculiar velocity --- regression distance'' relationship for
1327 RFGR galaxies. Galaxies of different samples are denoted by filled
symbols: circles --- sample ``1'', diamonds --- sample ``2'', squares ---
sample ``3'', triangles --- sample ``4'', crosses --- sample ``5''.}
\end{figure*}

This relation was again applied to the computation of peculiar velocities
of all 1327 galaxies. They are presented in Table\,2. The content of the
columns is as follows: \\
(1), (2) --- the number of the galaxy in the RFGC and FGC catalogues,
respectively; \\
(3) --- the right ascension and declination for the epoch 2000.0; \\
(4), (5) --- the corrected ``blue'' major and minor diameters, in arcmin; \\
(6) --- the corrected line width $W_c$ in km/s; \\
(7) --- the observed heliocentric radial velocity, in km/s; \\
(8) --- the radial velocity reduced to the centroid of the Local Group, in km/s;\\
(9) --- the radial velocity in the system of $3K$ cosmic microwave radiation,
in km/s; \\
(10) --- the distance (in km/s) measured from the basic regression on the
assumption that the model of motion of galaxies is dipole; \\
(11) --- the dipole component of the radial velocity, in km/s; \\
(12) --- the peculiar velocity, in km/s; \\
(13) --- the number of the sample from which the original data $V_h$ and
$W_{50}$ were taken.

Fig.\,3 shows $V_{pec}$ against $Hr$.

The relation between the predicted distance $Hr$ and the measured radial
velocity $V_{3K}$ is shown in Fig.\,4 for 1271 flat galaxies. The scatter
of points corresponds to $\sigma \approx 1200$\,km/s. After the exclusion
of 56 ``outliers'' the positive and negative radial velocities on the
``peculiar velocity --- distance'' relationship are located symmetrically
about zero and do not exhibit noticeable variation of slope with distance.

\begin{figure*}
\centerline{\psfig{figure=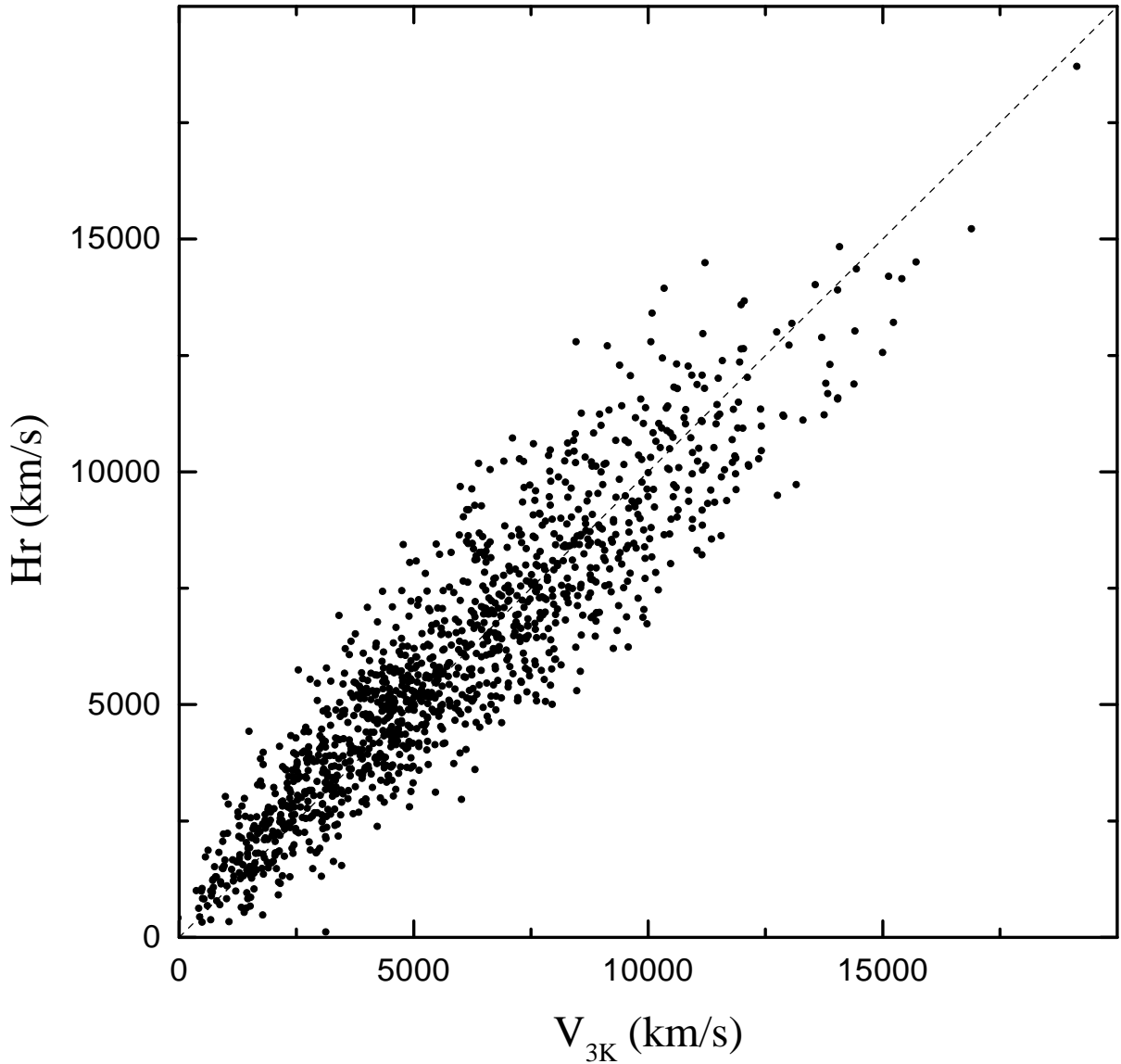,width=16cm}}
\caption{
The relationship between regression distance (in km/s) and radial velocity
for 1271 FRGC galaxies.}
\end{figure*}
\vspace{-0.2cm}
\section{Discussion and conclusions}
The problem of refinement of distances independent of their radial velocities
(in the case of spiral galaxies this is the improvement of the TF relation)
remains the most important, any samples being used. In many papers of the
past few years, necessity is shown of taking into account the
contribution made by both the surface brightness and the galaxy type to TF
relation.  Basing on detailed photometric investigation of spiral galaxies in
clusters, Willick (1999) would solve this problem in a more refined way. He
derived not strong but statistically significant dependence between TF relation
and surface brightness, and also between TF and galaxy light concentration index
``c''. In our case the value of $T$, which characterizes the variation of the
disk---bulge relationship along the Hubble sequence, may be a rough analog of
the index ``c''.

It seems promising to examine models of bulk motion of galaxies not only
in the most simple dipole approximation, but also with involvement of the
quadrupole (Willick et al., 1997a), as well as the quadrupole and octupole
(Parnovsky et al., 2000).

About 4\% of the galaxies from our list have considerable (over 3$\sigma$)
peculiar velocities. The existence of galaxies, moving at velocities
$\ga$(4000--5000)\,km/s with respect to the whole bulk, appears to be
fairly intriguing. We assumed that the cases of such deviations are
accounted for by the errors in measuring the basic observable quantities:
angular diameters, radial velocities, widths. These errors may be due to
morphology peculiarities, to the existence of a close neighbour, etc. After
the revision of these galaxies on the POSS-I and ESO prints we found no
discrepancies between values of newly measured diameters and the catalogue
ones. About half of the galaxies--``outliers'' have condensations near the
centre or along the disk (RFGC 59, 1396, 1441, 2637, 4080...), bent (RFGC
1033, 1272, 1719, 3396, 4054...), or strongly diffuse (RFGC 1234, 1719,
3200...) disks. Some galaxies with high peculiar velocities are located
in the region of clusters and have fairly close neighbours (RFGC 1542,
1568, 1897...). New observations will possibly elucidate the matter.

In conclusion the following will be noted: the Mark\,III catalogue and the
list presented here are currently the most complete, homogeneous and
covering all the sky samples for studying the field of peculiar velocities
of galaxies. Having a few common objects, they complement each other both in
types of galaxies and in depth of the survey.

Beginning with the creation of the first version of FGC, we have kept
searching for new estimates from literature and making observations of
radial velocities and widths of flat galaxies. In 1995, from the data for
about 800 FGC galaxies, we estimated the velocity modulus and the apex of
the bulk motion of flat galaxies, using the forward TF relation and the
dipole approximation: $|V|=260$\,km/s, $l=319\degr$, $b=28\degr$
(Karachentsev et al., 1995). At the second stage (Karachentsev et al., 2000)
a generalized multiparametric TF relation has been derived from which in a
dipole approximation over $\sim1000$ flat galaxies the values
$|V|=300$\,km/s, $l=328\degr$, $b=+7\degr$ have been obtained and an averaged
field of peculiar velocities has been constructed. Our data have turned out
to be in good agreement with the results obtained by Dekel et al. (1999),
$|V|=370$\,km/s, $l=305\degr$, $b=+14\degr$.

We have used the present list to investigate the bulk motion on the basis of the
generalized TF regression and involvement, in addition to the dipole component,
of the quadrupole and octupole components (Parnovsky et al., 2000). The bulk
motion direction in this case too is close to that obtained previously.

The data collected in this paper on radial velocities and widths of flat spiral
galaxies can be utilized as original when constructing the peculiar velocity
field by other methods.

\begin{acknowledgements}
The work was supported through the grant of the RFBR No.98-02-16100.
\end{acknowledgements}

{}

%\vspace{1cm}

%\clearpage
%\documentclass[11pt]{article}
%\usepackage[koi8-r]{inputenc}
%\usepackage[russian]{babel}
%\usepackage[saorus]{sao1}
%\usepackage{psfig}
%\newcommand{\E}{\rlap{E}}
%
%\begin{document}

\begin{onecolumn}
\begin{center}
\mbox{\hspace*{15cm}}\\
\topcaption{A list of velocity--distance data for the RFGC galaxies}
\tablehead{\hline
\multicolumn{1}{c}{RFGC}&
\multicolumn{1}{c}{FGC(E)}&
\multicolumn{1}{c}{RA (2000) D}&
\multicolumn{1}{c}{$a_o$}&
\multicolumn{1}{c}{$b_o$}&
\multicolumn{1}{c}{$W_c$}&
\multicolumn{1}{c}{$V_h$}&
\multicolumn{1}{c}{$V_{LG}$}&
\multicolumn{1}{c}{$V_{3K}$}&
\multicolumn{1}{c}{Hr}&
\multicolumn{1}{c}{$V_d$}&
\multicolumn{1}{c}{$V_p$}&
\multicolumn{1}{c}{$S$} \\ \hline}
\tabletail{\hline}
\par
\begin{supertabular}{rrrrrrrrrrrrr}
%\begin{tabular}{rrrrrrrrrrrrr}\\ \hline
   1& 2565&  000056$+$202017 & 2.02 & .17 & 416&  6804 & 7049 & 6452 &  6606 &  $-$206 &  52  & 1 \\
   4& 2567&  000136$+$033017 & 1.34 & .17 & 251&  6343 & 6531 & 5982 &  6355 &  $-$131 &$-$241& 1 \\
   6& 2563&  000204$+$803832 & 1.32 & .13 & 228&  4104 & 4362 & 4001 &  4735 &  $-$302 &$-$431& 2 \\
  12& 2573&  000223$+$271238 & 1.68 & .19 & 292&  7630 & 7891 & 7290 &  5499 &  $-$233 &2025  & 2 \\
  14&   1 &  000403$-$065845 &  .78 & .11 &  94&  3725 & 3868 & 3375 &  2426 &  $-$79  &1028  & 5 \\
  22&    6\E&000541$-$485848 &  .76 & .08& 120&  6612 & 6549 & 6411&  4507&  $ $140 &$  $1762 &5 \\
  27&    6 & 000653$+$414426 & 1.49 & .15& 254&  4247 & 4530 & 3951&  5125&  $-$278 &$-$895 &2 \\
  45&   16 & 001026$+$285917 & 1.98 & .16& 433&  7850 & 8111 & 7517&  7789&  $-$243 &$-$28 &1 \\
\end{supertabular}
%\end{tabular}
%\end{table}
\end{center}
%\end{document}
\end{onecolumn}

\end{document}